\documentclass[12pt]{article}

\usepackage{color}
\usepackage{amsmath}
\usepackage{amsfonts}
\usepackage{amssymb}
\usepackage{caption}
\usepackage{graphicx}
\usepackage{slashed}            
\usepackage{subfig}             
\usepackage{bbm}                
\usepackage{xspace}				
\usepackage{collref}			

\usepackage{tikz}
\usetikzlibrary{arrows,shapes}
\usetikzlibrary{trees}
\usetikzlibrary{matrix,arrows} 				
\usetikzlibrary{positioning}				
\usetikzlibrary{calc,through}				
\usetikzlibrary{decorations.pathreplacing}  
\usepackage{pgffor}							

\usetikzlibrary{decorations.pathmorphing}	
\usetikzlibrary{decorations.markings}
\tikzset{
    vector/.style={decorate, decoration={snake}, draw},
    fermion/.style={draw=black, postaction={decorate},
        decoration={markings,mark=at position .55 with {\arrow[draw=black]{>}}}},
    fermionbar/.style={draw=black, postaction={decorate},
        decoration={markings,mark=at position .55 with {\arrow[draw=black]{<}}}},
    fermionnoarrow/.style={draw=black},
    gluon/.style={decorate, draw=
        decoration={coil,amplitude=4pt, segment length=5pt}},
    scalar/.style={dashed,draw=black, postaction={decorate},
        decoration={markings,mark=at position .55 with {\arrow[draw=black]{>}}}},
    scalarbar/.style={dashed,draw=black, postaction={decorate},
        decoration={markings,mark=at position .55 with {\arrow[draw=black]{<}}}},
    scalarnoarrow/.style={dashed,draw=black},
	provector/.style={decorate, decoration={snake,amplitude=2.5pt}, draw},
	antivector/.style={decorate, decoration={snake,amplitude=-2.5pt}, draw},
}

\renewcommand{\tilde}{\widetilde} 

\newcommand{\arline}{\nonumber \\}
\newcommand{\dslash}[1]{\displaystyle{\not} #1}
\newcommand{\Tr}{\mathrm{Tr~}}

\newcommand{\beq}{\begin{equation}}
\newcommand{\eeq}{\end{equation}}

\usepackage[hypertexnames=false]{hyperref}		

\begin{document}
\begin{titlepage}

\vskip.5cm

\begin{center}
{\huge \bf A Top Seesaw on a 5D Playground}
\vskip.2cm
\end{center}

\begin{center}
{\bf {Don Bunk}$^a$, {Jay Hubisz}$^a$, {Jing Shao}$^a$, {Philip Tanedo}$^b$} \\
\end{center}
\vskip 8pt

\begin{center}
$^a$ {\it  Department of Physics, Syracuse University, Syracuse, NY  13244}

\vspace*{0.3cm}

$^{b}$ {\it Department of Physics, LEPP, Cornell University,\\ 142 Sciences Drive, Ithaca, NY 14853} \\

{\tt  
\href{mailto:djbunk@phy.syr.edu}{djbunk@phy.syr.edu}, \href{mailto:jhubisz@phy.syr.edu}{jhubisz@phy.syr.edu}, 
\href{mailto:jishao@syr.edu}{jishao@syr.edu}, 
\href{mailto:pt267@cornell.edu}{pt267@cornell.edu}}
\end{center}

\vglue 0.3truecm

\centerline{\large\bf Abstract}
\begin{quote}

We study quantum corrections to an extra dimensional Yukawa theory where a single flat extra spatial dimension is compactified on an interval.  At a UV scale this theory can be made equivalent to a 5D theory with a bulk four-fermion operator by choosing appropriate boundary conditions for the running of the low-energy theory.  Using the fermion bubble approximation we find vacuum solutions which break the chiral symmetries that arise from compactification.  Of particular interest are the form of brane localized terms which arise from fermion loops.  For example, quadratically divergent contributions to the scalar mass are absent at one loop due to a remnant of 5D Lorentz invariance that is only lost after introducing fermion bulk masses.  The model is interpreted as an implementation of top condensation in extra dimensions with an automatic seesaw mechanism.

\end{quote}

\end{titlepage}

\newpage

\tableofcontents


\section{Introduction}
\label{intro}

Models of top-quark condensation~\cite{Miransky:1988xi,Miransky:1989ds,Marciano:1989mj,Marciano:1989xd,Bardeen:1989ds} are particularly appealing models of electroweak symmetry breaking.  These theories are relatively compact and have the feature of automatically generating a large top Yukawa interaction with a composite Higgs field that is a bound state of a top--anti-top pair.  While the simplest model is plagued by naturalness issues, subsequent embeddings of top condensation in supersymmetric~\cite{Carena:1991ky} and strongly coupled models of electroweak symmetry breaking (EWSB)~\cite{Eichten:1979ah,Hill:1991at,Hill:1994hp} can reproduce the weak scale without excessive fine tuning.  However, a combination of flavor~\cite{Buras:1982ff,Buchalla:1995dp,Burdman:2000in,Simmons:2001va} and electroweak precision constraints~\cite{Peskin:1990zt,Peskin:1991sw} have consistently put tension on implementations of top condensation within strongly coupled scenarios.  For a review with extensive discussion of these issues and a complete citation list, see~\cite{Hill:2002ap}.

Recent focus on extra dimensional models of EWSB, particularly those constructed on geometrically warped backgrounds~\cite{Randall:1999ee,Randall:1999vf} has shed new light on naturalness issues of the electroweak sector, and how precision tests might be addressed in a weakly coupled framework.  In this paper, we explore the possibility of embedding top condensation within an extra dimensional setup.  Such models with warped geometry are expected to generate natural hierarchies of scales.   In this paper, we explore the 5D Nambu--Jona-Lasinio (NJL) mechanism~\cite{Nambu:1961fr, Nambu:1961tp} in a flat space toy model, with the idea that many of the results will carry over to more realistic extra dimensional scenarios utilizing a warped compactification.

In calculating the 5D effective action for fermion--antifermion bound states, we renormalize a 5D Yukawa theory compactified on an interval.  The running is supplemented by an ultraviolet (UV) ``composite" boundary condition at a scale $\Lambda_0$.  At the UV boundary, which we take to be at an energy greater than the compactification scale $1/L$, the theory describes 5D fermions that interact via a four-fermion interaction which arises from unspecified UV dynamics, perhaps from physics above the cutoff due to the strong coupling limit of the extra dimensional model.  In deconstruction models~\cite{ArkaniHamed:2001ca,Hill:2000mu}, where the extra dimension resolves into a product gauge structure at high energies, the four-fermion operator could arise as a result of the (unspecified) dynamics which breaks the product group structure down to the Standard Model (SM) at low energies.   The four-fermion operator could also arise due to intrinsically 5D dynamics such as a spontaneously broken 5D gauge theory.  

Top condensation has been studied in extra dimensional contexts previously~\cite{Dobrescu:1998dg,Cheng:1999bg,Rius:2001dd,Burdman:2007sx,Bai:2008gm}, although focus has typically been on the low-energy theory below the scale of compactification.  Our analysis includes the effect of 5D running up to the scale associated with the four-fermion interaction, and gives predictions for a Kaluza-Klein (KK) tower of scalar bound states corresponding to a 5D composite field.  Of particular interest are the form of and role played by brane localized terms generated by fermion loops.  Other top condensation models that simultaneously generate the correct top and $W$-boson masses generally supplement top condensation with a seesaw mechanism~\cite{Dobrescu:1997nm,Chivukula:1998wd}.  Features of our 5D construction are similar to those found in extra dimensional top see-saw models~\cite{Cheng:2001nh,He:2001fz}, in which the lightest KK excitations of the fermions play a key role in the formation of the condensate.

We begin with a review of the 4D NJL model, which we then extend to a 5D setup compactified on an interval, or equivalently, an $S_1/\mathbb{Z}_2$ orbifold.  Extension of these methods to a compactified model is relatively straightforward, although there are some complications associated with performing quantum corrections in an extra dimensional model, which we discuss.  We work in the fermion bubble approximation, valid as long as the scale associated with the four-fermion operator is below the scale at which any additional 5D interactions (i.e. gauge interactions of the SM) become strongly coupled.  Section~\ref{sec:5Dloops} contains a study of the relevant fermion loop graphs in 5D flat space.   We then calculate the 5D quantum effective action valid at low scales.  Solving the scalar equations of motion in this effective theory determines whether or not a chiral symmetry breaking condensate is formed.  

We calculate the resulting light fermion and scalar spectrum, requiring a weakly gauged $SU(2)_L \times U(1)_Y$ version of the model to reproduce the observed $W$-boson mass.  We find that the top quark mass and $W$ mass constraints can be simultaneously satisfied by making an appropriate choice of the fermion bulk mass parameters.  The lowest lying scalar fluctuation is found to be generically heavy, due primarily to a large effective quartic coupling generated in the model.  Lighter values can be generated by going to larger $N_c$ or creating a larger hierarchy between the four-fermion scale and the compactification scale $\Lambda_0 L \gg 1$.  The second of these choices is made at the expense of increased fine-tuning of the interaction strength associated with the four-fermion operator and reducing the validity of the fermion bubble approximation.

\section{Extending the NJL Model to 5D}
\label{sec:5DNJL}
A toy model for spontaneous breaking of chiral symmetry in four dimensions can be constructed with a low-energy effective theory of massless fermions supplemented with a single chirally symmetric four-fermion contact operator~\cite{Nambu:1961fr, Nambu:1961tp}.  The Lagrangian for this model, valid at the scale $\Lambda$ is
\begin{equation}
\label{eq:4DNJLnophi}
\mathcal{L} = \bar{\psi} i \displaystyle{\not} \partial \psi + \frac{g^2}{4\Lambda^2} \left[ (\bar{\psi} \psi)^2-  (\bar{\psi} \gamma^5 \psi)^2   \right]
\end{equation}
where $\psi$ is a 4-component massless Dirac fermion.  The Lagrangian is invariant under independent chiral rotations of the left- and right-handed components of $\psi$.

In two component notation, utilizing a complex auxiliary scalar field $\phi$, we can re-write this Lagrangian as
\begin{equation}
\mathcal{L} = \bar{\psi}_L i \displaystyle{\not} \partial \psi_L +\bar{\psi}_R i \displaystyle{\not} \partial \psi_R + g \phi \bar{\psi}_L \psi_R + \mathrm{h.c.} - \Lambda^2 |\phi |^2.
\label{eq:4DNJL}
\end{equation}
The field $\phi$ carries chiral charge such that this Lagrangian has the same symmetry as Eq.~(\ref{eq:4DNJLnophi}).
Running down this theory from the scale $\Lambda$ to a low scale $\mu$, taking into account only fermion loops, one finds that the scalar field $\phi$ develops dynamics and a quartic interaction.  The fermion loop contribution to the scalar mass$^2$ is negative, and for sufficiently strong coupling, $g$, the quantum corrections overcome the positive $\Lambda^2 | \phi |^2$ term.  In this case, the scalar field then picks a vacuum expectation value (vev), and breaks the chiral symmetry of the theory.  

This mechanism was posited as a method to spontaneously break the electro-weak gauge interactions, where the fermion bound state consisted of top/anti-top pairs~\cite{Bardeen:1989ds}.  A particularly appealing feature of this construction is the presence of a quasi-infrared fixed point in the top Yukawa coupling which renders the top Yukawa relatively insensitive to the compositeness scale~\cite{Pendleton:1980as,Hill:1980sq}.    Above this fixed point, the top Yukawa blows up in the UV, and the coupling is in the domain of attraction for this fixed point which resides at a value of $\lambda_t \sim 1$.

We consider a 5D version of the above model, in which there is a four-fermion operator that leads to a composite five dimensional scalar field.  This operator must arise from some UV dynamics, as in the case of 4D top condensation models~\cite{Hill:1991at}. In this work, we do not specify this dynamics and focus on the mechanics of the renormalization of this theory.  A model with better UV behavior is currently under investigation.

The theory at a high scale $\Lambda_0$ consists of two 5D Dirac fermions, $\Psi_L$ which contains a left-handed zero mode  in the spectrum, and $\Psi_R$ which contains a right handed one.  Other assignments are possible, and will have different IR structure, however this theory is the one that most easily generalizes to a standard model-like low-energy spectrum.  In addition, the chiral symmetries of this model are identical to those in Eq.~(\ref{eq:4DNJL}).    We write the action for the theory at the scale $\Lambda > 1/L$ as defined on a circle with perimeter $2L$:
\begin{equation}
S_\text{5D NJL} = \int d^4x \int_{-L}^L dz \bar{\Psi}_L \left(i \displaystyle{\not} \partial -M_L(z) \right) \Psi_L +\bar{\Psi}_R\left( i \displaystyle{\not} \partial - M_R (z) \right) \Psi_R+ \frac{g^2}{\Lambda_0^3} \bar{\Psi}_L \Psi_R \bar{\Psi}_R \Psi_L.
\label{eq:5DNJL}
\end{equation}
where $\displaystyle{\not} \partial \equiv \gamma^\mu \partial_\mu + i \gamma^5 \partial_z$ and all fields are assigned periodic boundary conditions.  The spectrum of the theory is then reduced by performing the identification $z \leftrightarrow -z$ which restricts the physical region of the space to the interval $z \in [0,L]$.  The field solutions that remain can be either odd or even under this identification, although all operators in the Lagrangian must be even.  The orbifold assignments that produce the spectrum described above are:
\begin{equation}
\Psi_L (z) = -\gamma^5 \Psi_L (-z) \text{, and } \Psi_R (z) = \gamma^5 \Psi_R (-z).
\end{equation}
In order for the action to be invariant, the fermion mass terms must be odd under the orbifold assignment:  $M_{L,R} (z) = - M_{L,R} (-z)$.

While this procedure is equivalent to beginning with an interval and assigning boundary conditions~\cite{Csaki:2003sh,Csaki:2005vy}, we show in Appendix~\ref{app:5Dtranslation} that the orbifold language allows a simple, intuitive explanation for the presence or lack of certain brane localized terms that are induced by quantum corrections.

We assume mass profiles which are constant in the physical region, discontinuously jumping at the orbifold boundaries to satisfy the boundary condition above:
\begin{equation}
M_{L,R} (z) = \left\{ \begin{array}{cl} +m_{L,R} & z > 0 \\ - m_{L,R} & z < 0. \end{array} \right.
\label{eq:fermmasses}
\end{equation}
The zero modes are then exponentially localized, with profiles given by:
\begin{eqnarray}
\Psi_L^0 (x;z) = \sqrt{\frac{m_L}{1-e^{-2 m_L L}} } e^{- m_L |z|} \nonumber \\
\Psi_R^0 (x;z) = \sqrt{\frac{m_R}{e^{2 m_R L}-1} } e^{ m_R |z|}.
\end{eqnarray}
In the 4D low-energy effective theory and ignoring quantum effects, the zero modes couple via a four-fermion operator that has a form identical to that of Eq.~(\ref{eq:4DNJL}), with effective four-fermion coupling given by an overlap of the zero mode wave functions: 
\begin{equation}
\frac{g_{4D}^2}{\Lambda_\text{eff}^2} = \frac{g^2}{\Lambda_0^3} \frac{m_L m_R}{m_L-m_R} \left( \coth m_L L - \coth m_R L \right),
\end{equation}
which is exponentially suppressed in the case that both $m_L$ and $m_R$ are the same sign, and the LH and RH zero modes are localized on opposite boundaries of the physical region.  We will show that scalar bound states and chiral symmetry breaking with scales well below the scale $1/L$  can still be obtained, regardless of this suppression.  In the KK mode interpretation, these scalars are presumably relativistic deeply bound states of a combination of KK modes.  This strongly suggests that a full 5D calculation including all KK modes below the cutoff $\Lambda_0$ should be performed in order to properly formulate the low-energy theory.

  To analyze the IR behavior of this theory, we write the 5D four-fermion interaction in terms of a complex auxiliary field $\phi$.  At the scale $\Lambda_0$, the theory is then a model of Yukawa interactions in which the scalar field has no dynamics: 
\begin{align}
S_\text{5D NJL} = \int d^4x \int_{-L}^L dz &\bar{\Psi}_L \left( i \displaystyle{\not} \partial -M_L(z) \right) \Psi_L +\bar{\Psi}_R \left( i \displaystyle{\not} \partial  - M_R(z) \right) \Psi_R \nonumber \\
& - \Lambda_0^2 | \phi |^2 + \frac{g}{\sqrt{\Lambda_0}} \phi \bar{\Psi}_L \Psi_R + \text{h.c.} \label{eq:auxL}
\end{align}
Integrating out the field $\phi$ reduces Eq.~(\ref{eq:auxL}) to Eq.~(\ref{eq:5DNJL}).  The main calculation of this paper will be on running this effective Lagrangian down to a low scale $\mu < \frac{1}{L}$, and solving the low-energy equations of motion for the scalar field.  We calculate the running in the ``fermion bubble" approximation, integrating out only the fermionic contribution to the scalar effective action.  This approximation is the analog of re-summing the fermion ladder diagrams in the theory written down in Eq.~\ref{eq:5DNJL}.


\section{Quantum Corrections in 5D}
\label{sec:5Dloops}
In models with compactified extra dimensions, quantum corrections are complicated by the fact that momenta along the compactified directions are discrete while the 4D momenta span a continuum.  In our model, momenta along the compactified coordinate are quantized in units of $n \pi/L$, where $L$ is the size of the physical region.  In this section, we compute these quantum corrections for the Yukawa theory in Eq.~(\ref{eq:auxL}).

Quantum effects in extra dimensional models have been studied in some contexts, particularly for the running of gauge couplings~\cite{Lewandowski:2002rf,Lewandowski:2004yr,Contino:2002kc}.  Such calculations are often made simpler due to gauge invariance, which ensures that calculating the running of the coupling of the zero mode gauge field, which has a constant extra dimensional profile, is sufficient to describe all running effects in 5D.  Our analysis of a 5D Yukawa theory must be intrinsically five dimensional, taking into account all possible external scalar states, since there is no such underlying symmetry which keeps the lowest lying mode flat.

In determining the quantum effects of the 5D theory, there is the approach of determining the KK spectrum, integrating out the extra dimension, and then truncating the effects of the tower at the desired level of accuracy.  It is then a matter of computing usual 4D Feynman diagrams using these few KK modes.  This approach, however, obscures 5D translation invariance, and is in fact quite complicated if more than a couple KK modes are included.  This is especially the case in this construction, since there are a large number of possible scalar bound states.  When the equation of motion is applied on the scalar field $\phi$ at the scale $\Lambda_0$, and the fermions are expanded in terms of their KK towers, we find:
\begin{equation}
\phi =\frac{g}{\Lambda_0^{5/2}} \bar{\Psi}_R \Psi_L = \frac{g}{\Lambda_0^{5/2}}  \sum_{m,n} \bar{\psi}^m_R \psi^n_L.
\end{equation} 
Quantum effects below the scale $\Lambda_0$ mix these fermion bi-linears with each other, and the effective action must then be re-diagonalized.  It is much simpler and perhaps more illuminating to instead compute all quantum effects from the 5D viewpoint, and then solve the resulting 5D scalar equation of motion.

The most straightforward method is to compute all quantum corrections in momentum space, where the effects of orbifolding are taken into account in the form of the propagators.  Either a hard momentum cutoff or dimensional regularization may then be used to study the divergence structure of the theory.  The first of these is most suited to the 5D NJL model, since it explicitly contains information about power law divergences.  Dimensional regularization, on the other hand, automatically subtracts these, leaving only poles corresponding to logarithmic divergences.  We study both regulators, the former because it applies well to models with an explicit cutoff, and the latter since it is a point of interest to see how the 5D divergence structure, which contains no bulk log divergences, is obtained from the 4D KK tower which contains an infinite number of them.

It is, in principle, possible to use a mixed position-momentum space basis, where the propagators depend on the position in the extra dimensional coordinate, however in this case it is unclear how one would implement a regularization procedure which respects local 5D Lorentz invariance.


\subsection{Quantum corrections with vanishing fermion bulk masses}

In the case that the bulk fermion masses vanish, the fermion propagators are not difficult to compute.  The Yukawa theory under consideration is then similar to the one examined in~\cite{Georgi:2000ks}, but with slightly different orbifold assignments and field content.  In this section we utilize the notation of these authors.  In particular, a derivation of the fermion propagators can be found in Section 2 of that publication.

In 5D momentum space, the fermion propagators are given by:
\begin{equation}
S_F^{(L, R)} (p ; p_5 , p'_5) = (2 L) \frac{i}{2}   \left\{ \frac{  \delta_{p_5, p'_5}}{\displaystyle{\not} p + i \gamma^5 p_5}  \pm \frac{ \delta_{-p_5, p'_5}}{\displaystyle{\not} p + i \gamma^5 p_5} \gamma^5 \right\}
\end{equation} 
where the $+$ is for a 5D fermion in which a left-handed zero mode survives the orbifold projection, and the $-$ is for a 5D fermion which contains a right-handed zero mode in the spectrum\footnote{We have chosen a convention in which the period of the Fourier series appears in the Kronecker-$\delta$s of momentum ($2 L~\delta_{p_5,k_5}$), and in sums over unconstrained 5D momenta ($\frac{1}{2L} \sum_{k_5}$).  This makes it simpler to compare with the (mostly) standard treatment in non-compact dimensions where the transformation to momentum space comes with a $\frac{1}{2 \pi}$ normalization.   The dictionary between the compact and non-compact 5D theory consists of replacing sums with integrals, Kronecker-$\delta$s with $\delta$-functions, and all factors of $2L$ with $2 \pi$.}.   The 5D momentum is given by $p_5 = \frac{n \pi}{L}$, where $n$ ranges over all integers.  The fermion propagators conserve the magnitude of the 5D momentum, but only up to a sign.  The breaking of 5D translation invariance is a manifestation of the reflection conditions at the orbifold fixed points.  The remaining conservation of KK number is a tree level symmetry of the theory that is present in the limit of vanishing bulk mass.

We are interested in computing the scalar two- and four-point functions.  Since interaction terms in extra dimensional theories are non-renormalizable, higher dimensional operators will be generated as well.  For the purposes of illustration in this toy model, we ignore these contributions.  One could, in principle, arrange for these terms to be removed via fine tuning of the coefficients of such operators against the quantum corrections to them.  This tuning should then presumably be derived as a natural consequence of some UV complete model.


\begin{center}
{\bf The scalar two-point function}
\end{center}

\begin{figure}[h]
\center{\includegraphics[width=4in]{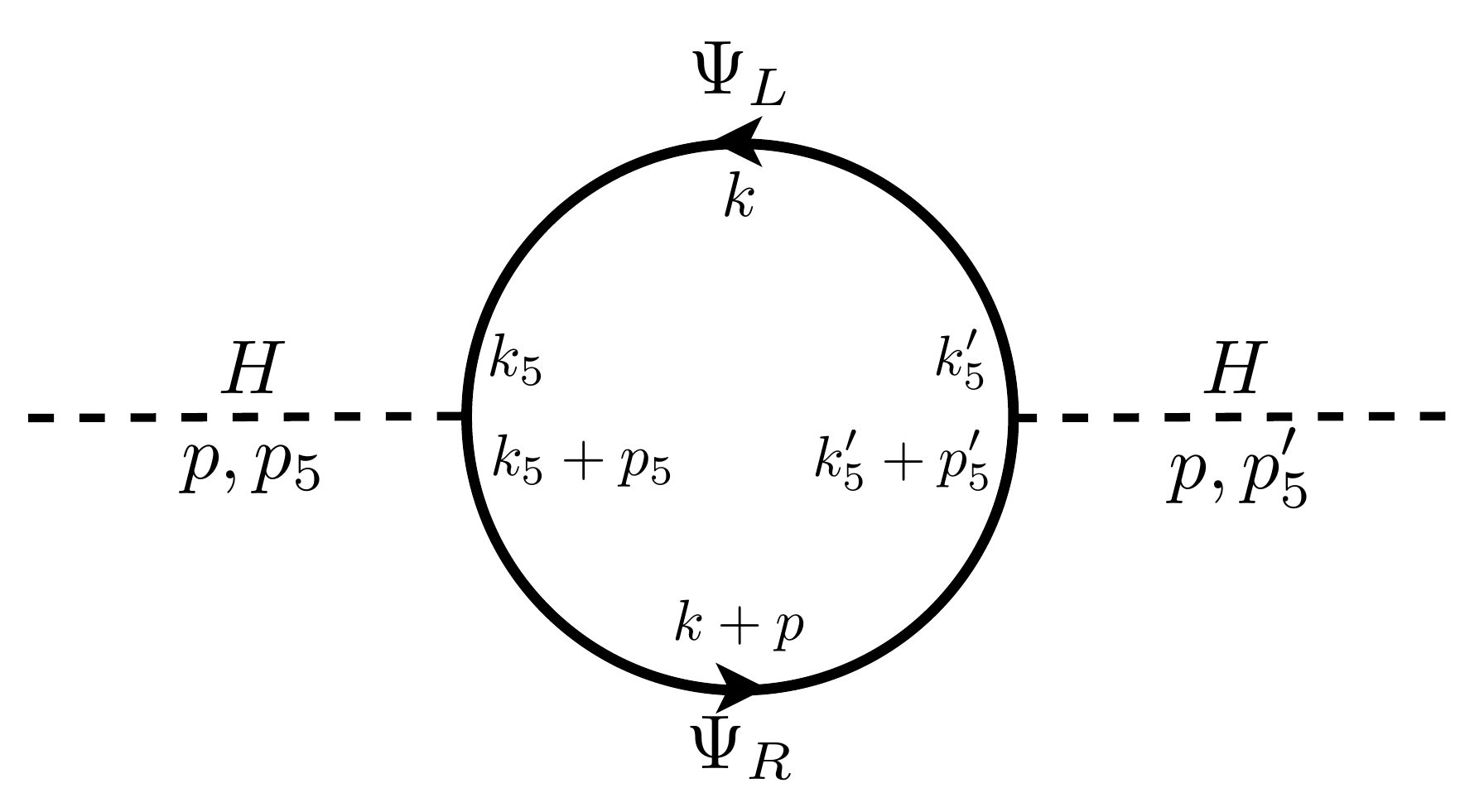}}
\caption{The 5D scalar two point function, where the scalar couples to two flavors of 5D Dirac fields, each of which contains either LH and RH zero mode in the KK mode spectrum.}
\label{fig:twopoint}
\end{figure}

In the massless fermion bubble approximation, the scalar two-point function at one loop consists of the diagram shown in Figure~\ref{fig:twopoint}.   In the compactified 5D theory, this single diagram encapsulates the quantum corrections to the bulk kinetic and mass terms.  In addition, it also contains information about brane localized terms which are quadratic in the scalar field.  This diagram gives information about how to run the scalar sector of the Yukawa theory from the high scale $\Lambda_0$ down to low energies. 
The value for the diagram is
\begin{eqnarray}
&&- \frac{g^2}{\Lambda_0} \sum_{k_5, k'_5} \int \frac{d^d k}{(2\pi)^d}\Tr\left[ \frac{ (\slashed{k} + i \gamma^5 k_5 ) ( \delta_{k_5, k'_5} - \gamma^5 \delta_{k_5, -k'_5} ) }{k^2 - k_5^2} \right. \arline
&& \left. \cdot  \frac{ (\slashed{k}+\slashed{p} + i \gamma^5 [ k'_5 +p_5' ] ) ( \delta_{k_5+p_5, k'_5+p'_5} + \gamma^5 \delta_{k_5+p_5, -k'_5-p'_5} ) }{(k+p)^2 - (k'_5+p'_5)^2} \right].
\end{eqnarray}
 
  Let us first discuss brane localized divergences of the two-point diagram.   In extra dimensional theories, it is now well known that quantum effects generally violate KK-number conservation~\cite{Georgi:2000ks,Cheng:2002iz,Carena:2002me}.   The presence of brane localized terms can be identified by divergences which do not conserve 5D momenta.  Such divergences signal that a counterterm is necessary, and that the brane term should be included in the tree level action.  Expanding the numerator of the diagram and simplifying the Kronecker-$\delta$s, there are in principle terms proportional to $\delta_{p_5, p_5'}$, $\delta_{-p_5, p_5'}$,  $\delta_{2 k_5, -p_5-p'_5}$, and $\delta_{2 k_5, p'_5-p'_5}$.  The first two types of terms conserve 5D momentum up to a sign and hence correspond to bulk corrections, while the second two Fourier transform into $\delta$-functions at the brane positions and so correspond to brane localized terms.
   
 Applying the usual Dirac trace identities, the brane localized terms vanish.  This is perhaps somewhat surprising at first glance.  One might expect that there are brane localized quadratic divergences which renormalize the scalar mass independently on the branes versus in the bulk.  One might also expect the generation of brane localized kinetic terms for the scalar field.  The reason for the absence of such terms at the one-loop level is that 5D translation invariance is not broken severely enough in this process, as explained in Appendix~\ref{app:5Dtranslation}.  In fact, there are a variety of scenarios in which brane localized terms are not generated at the one-loop level. 
  
Let us now identify the bulk renormalization terms.  We expect a cubically divergent mass renormalization, and a linear divergence in the 5D kinetic terms.  One of the bulk renormalization terms is proportional to $\delta_{p_5,p_5'}$, the other  $\delta_{p_5,-p_5'}$ (effectively reflected and transmitted waves through the orbifold fixed points).
From the trace, these have the following momentum structure:
\begin{equation}
\frac{k \cdot (k+p) - k_5 (k_5+p_5')}{\left(k^2-k_5^2 \right) \left( (k+p)^2 - (k_5+p_5')^2 \right)}
\end{equation}
The $k_5$ are quantized on $k_5 = n \pi/L$, with $n$ any integer.  This means that the 5D sum cannot be shifted, while the 4D momenta can be redefined in the usual way in order to make the Wick rotated integrand spherically symmetric in Euclidean momentum.

The coefficients of the $\delta_{p_5, p'_5}$ and $\delta_{p_5, -p'_5}$ terms are identical.  After combining denominators using Feynman parameters, they are given by:
\begin{equation}
- \frac{g^2}{4 \Lambda_0} \sum_{k_5} \int \frac{d^d k}{(2\pi)^d} \int_0^1 dx \frac{(l^2 - l_5^2)-x (1-x) (p^2-p_5^2) +l_5 p_5 (2x-1)}{[(l^2-l_5^2) + x (1-x) (p^2-p_5^2)]^2},
\end{equation}
where $l_5 = k_5+ x p_5$.  Unfortunately, one cannot shift the 5D momentum in the sum this way since $l_5$ is not quantized on the same spectrum as $k_5$ and the above expression is only a heuristic presentation.

This lack of shift invariance highlights the fact that a naive hard cutoff  for the 4D momentum integrals obscures the underlying physics.  Such a procedure explicitly violates 5D Lorentz invariance, and will lead to apparent violation of the spacetime symmetries by short-distance interactions.  For example, if one performs the sum over \emph{all} unconstrained five-momenta, one obtains an analytic expression as a function of the 4D loop momentum.  The remaining integrand can then be performed with a hard cutoff, expanded in small external momenta, and then interpreted as a contribution to the effective action.  The resulting expression contains terms proportional to $p_\mu^2$ and $p_5^2$ with coefficients which differ in general.  5D Lorentz invariance can then be restored by fine tuning separate counter terms order by order in perturbation theory, but the connection with the original 5D theory defined at the physical scale $\Lambda_0$ is then lost.  To properly formulate the low energy dynamics, one must choose the regulator more carefully.  

We first perform the integration utilizing dimensional regularization.  Since there is no explicit cutoff scale, there are no subtleties about the regularization procedure respecting local 5D Lorentz invariance.  Performing the 4D momentum integration first, we have
\begin{equation}
i \Pi (p^2, p_5) = - i \frac{g^2}{4 \Lambda_0} \sum_{k_5} \int_0^1 dx \frac{\Delta^{d/2-2}}{(4\pi)^{d/2}} \left\{ \frac{d}{2} \Delta \Gamma (1-d/2) + \left[ x (1-x) p^2 + k_5^2+p_5 k_5 \right] \Gamma (2-d/2) \right\}
\end{equation}
where $\Delta$ is given by:
\begin{equation}
\Delta = - x (1-x) (p^2 - p_5^2) + (k_5 + x p_5)^2.
\end{equation}  
Using zeta-function regularization for the remaining sum over 5D internal loop momentum we have
\begin{equation}
i \Pi(p^2, p_5^2) = i\Pi (0) -\frac{i g^2}{8 \Lambda (4 \pi)^{d/2}} \left(\frac{\pi}{L} \right)^{4-d} \left(2 \zeta (4-d)+(\mu_\text{IR} L)^{d-4} \right) \Gamma (2-d/2) \left[ p^2 + p_5^2 \left(  2- d \right) \right]. 
\end{equation}
We have regulated the contribution of the zero mode with an IR cutoff, $\mu_\text{IR}$.  The two point function for vanishing external momentum, $i\Pi(0)$, is given by:
\begin{equation}
i \Pi(0) = - i \frac{g^2}{4 \Lambda (4 \pi)^{d/2}} \left(\frac{ \pi}{L} \right)^{d-2} \zeta (2-d) \Gamma (1-d/2)
\end{equation}
Taking the limit as $d \rightarrow 4$, with $\epsilon_\text{IR} \equiv \mu_\text{IR} L$, we have the final result:
\begin{equation}
i \Pi(p^2, p_5^2) = \frac{i g^2}{4 \Lambda (4 \pi)^{2}} \left[ 2 \left(\frac{ \pi}{L} \right)^{2} \zeta'(-2) + \log (2 \pi \epsilon_\text{IR}) \left( p^2 - 2 p_5^2 \right) \right]
\end{equation}

Let us point out some aspects of these results:  First, all expressions are finite as $d \rightarrow 4$.  For the field strength term, the pole in the $\Gamma$ function is canceled by the sum of the zeta function and the contribution of the zero mode.  That is, the UV divergences created by the zero mode are canceled by the UV divergences of the tower of KK modes.  Second, note that the coefficient of the $p^2$ and $p_5^2$ terms differ in the limit $d \rightarrow 4$.  These finite terms correspond to non-local contributions to violations of 5D translation invariance from the presence of the orbifold fixed points.

The finiteness of the result in this regularization scheme is expected.  Since all divergences must be local, the UV structure of the bulk compactified theory should match that of the uncompactified model.  All divergences in noncompact odd dimensions are power laws and are automatically subtracted when using dimensional regularization. So both the compact and uncompact models yield finite results for the two-point function in this regularization scheme.

It is possible to utilize a hard cutoff regularization scheme which respects the local spacetime symmetries.  This is beneficial, since such a scheme has a better physical interpretation in terms of our physical cutoff, $\Lambda_0$.  The procedure is described in detail in Appendix~\ref{app:euler-maclaurin}, but in many cases it consists simply of approximating the sum over momenta by an integral, at which point the integrand is manifestly 5D Lorentz invariant, and integration over the interior of a four-sphere in the loop momentum can be performed in the standard way.  The substitution required is $\frac{1}{2L} \sum_{k_5} \rightarrow \int \frac{dk_5}{2 \pi}$.
 The two-point function in this regularization scheme is then
 \begin{equation}
 i \Pi (p^2; p_5,p_5') =  \left( \delta_{p_5,p_5'} + \delta_{p_5,-p_5'} \right)  \frac{g^2L}{2 \Lambda_0} \int \frac{d^5k}{(2 \pi)^5}  \int_0^1 dx \frac{(l^2 - l_5^2)-x (1-x) (p^2-p_5^2) +l_5 p_5 (2x-1)}{[(l^2-l_5^2) + x (1-x) (p^2-p_5^2)]^2},
 \end{equation} 
 and we can now shift the full 5D loop momentum in the usual way, and use a 5D hard cutoff $\Lambda$.  The result, as an expansion in $P^2 = p^2 - p_5^2$, is given by
 \begin{equation}
 i \Pi (p^2; p_5,p_5') = i L \left( \delta_{p_5,p_5'} + \delta_{p_5,-p_5'} \right) \left[ \frac{g^2 \Lambda^3}{18 \pi^3 \Lambda_0} + \frac{g^2 \Lambda}{10 \pi^3 \Lambda_0} P^2 \right]
 \equiv L \left( \delta_{p_5,p_5'} + \delta_{p_5,-p_5'} \right) i \tilde{\Pi} (P^2).
 \label{eq:2pthardcutoff}
\end{equation}
We have kept $\Lambda_0$ separate from the regulator cutoff in this expression to highlight the sensitivity to an arbitrary UV scale, although we take them to be equal in our final expression for the effective action.  Implicit in Eq.~(\ref{eq:2pthardcutoff}) is an IR scale, $\mu \ll \Lambda$, which can be put into the effective action with the replacements $\Lambda^n \rightarrow \Lambda_0^n - \mu^n$.

\begin{center}
{\bf The scalar four-point function}
\end{center}

\begin{figure}[h]
\center{\includegraphics[width=4in]{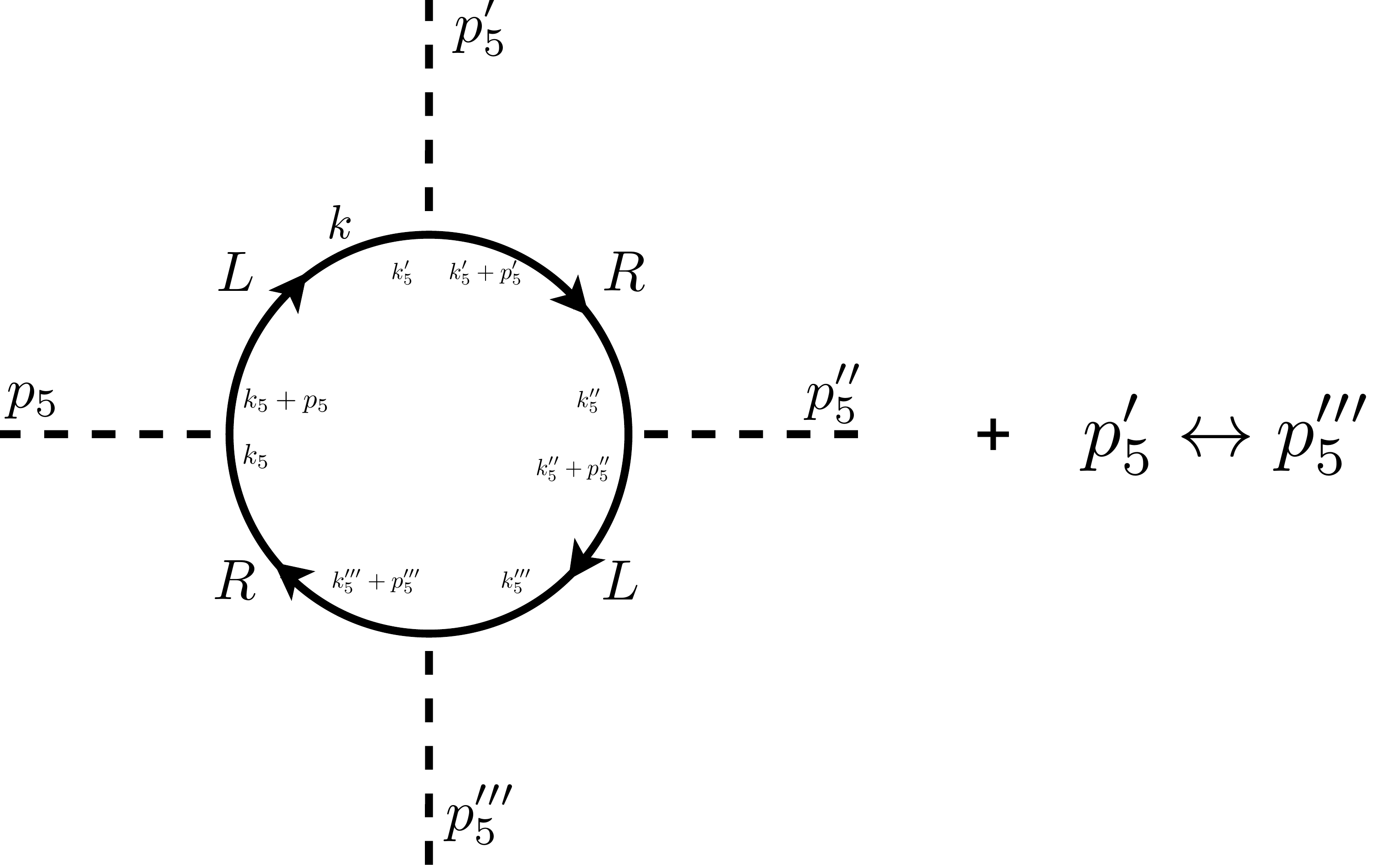}}
\caption{The 5D scalar four point function.}
\label{fig:fourpoint}
\end{figure}

The quartic coupling also renormalizes, although we again find that all divergences are confined to the bulk.  The relevant Feynman diagrams are shown in Figure~\ref{fig:fourpoint}, and evaluate to
\begin{align}
i V_4(0;p_5,p'_5,p''_5,p'''_5) = - \frac{g^4}{\Lambda_0^2} \sum_{ \substack{ k_5, k'_5,  \\k''_5,k'''_5}} \int \frac{d^4k}{(2\pi)^4} \Tr &\left[S^{R}_F(k;k_5,k_5'''+ p_5''') S^{L}_F(k;k_5''',k''_5 + p''_5) \right. \nonumber \\ 
& \times \left.  S^{R}_F (k;k_5'',k_5' + p_5') S^L_F (k;k_5''',k_5'' + p_5'')   \right].
\end{align}
Terms which contribute to bulk running of the quartic arise from an even number of insertions of the 5D momentum conserving Kronecker-$\delta$s while terms which contribute to brane running of the quartic involve an odd number of these.  The potential brane terms each involve (at leading order in loop momenta) the trace of four identical Dirac matrices, $\slashed{k}$, with a $\gamma^5$, and therefore vanish.

Performing the calculation using dimensional regularization again produces a finite result, with KK modes canceling against the contribution of the zero modes.  We only present the result utilizing a 5D Lorentz invariant hard cutoff.  
We find
\begin{align}
i V_4(0;p_5,p'_5,p''_5,p'''_5) =  \frac{- i g^4 \Lambda}{24 \pi^3 \Lambda_0^2}    (2L) \sum_{\pm} \delta_{ 0, p_5  \pm p'_5 \pm p''_5  \pm p'''_5}.
\end{align}
Where the sum is over all 8 permutations of signs in the Kronecker-$\delta$.

To summarize the results of this section, we find that the bulk UV structure of the theory is as expected, where the running is purely power law.  We have explicitly shown the cancellation of log divergences in the dimensional regularization scheme for the two-point function.  

The one-loop brane localized divergence structure is different from naive expectations.  Despite the intuition that brane localized terms should be forced by breaking translation invariance via the orbifold identification, they are not generated at one loop.  As we discuss in Appendix~\ref{app:5Dtranslation}, this is due to the interplay of the left- and right-handed components of 5D fermions.

\subsection{Quantum corrections with fermion bulk masses}

The arguments that protect against brane localized terms fail when fermion mass terms are added into the theory.  Under the orbifolding procedure, such masses must be odd under the projection since the fermion bilinears $\bar{\Psi} \Psi$ are odd.  These masses could arise from a scalar domain wall to which the fermions are coupled via a Yukawa interaction.  These domain walls are trapped at the orbifold fixed point by the orbifold quantum numbers of this scalar field and give rise to fermion localization in the extra dimension~\cite{Kaplan:1992bt,Mirabelli:1999ks,Kaplan:2001ga}.  Because such fermion masses explicitly break 5D translation invariance at the orbifold fixed points, it is expected that they generate brane localized terms.

In this section, we calculate the quantum corrections in the presence of fermion bulk masses.  These mass terms do not conserve even the magnitude of the 5D momenta so that the explicit form of the propagators in momentum space is rather complicated to compute.  However, we can accurately capture the divergence structure of the theory by treating the 5D mass term as a perturbation to the massless scenario.  

We take the fermion masses to have the profiles given in Eq.~(\ref{eq:fermmasses}).  To obtain the Feynman rule in momentum space, we compute the Fourier series of the fermion mass terms in the action, and read off the interaction vertex.  Since the mass term switches sign at the orbifold fixed points, its Fourier series is non-trivial.  That is, the mass term acts as a source for 5D momentum which can be injected into a given diagram.  The Feynman rule is:
\begin{equation}
\includegraphics[width=3in]{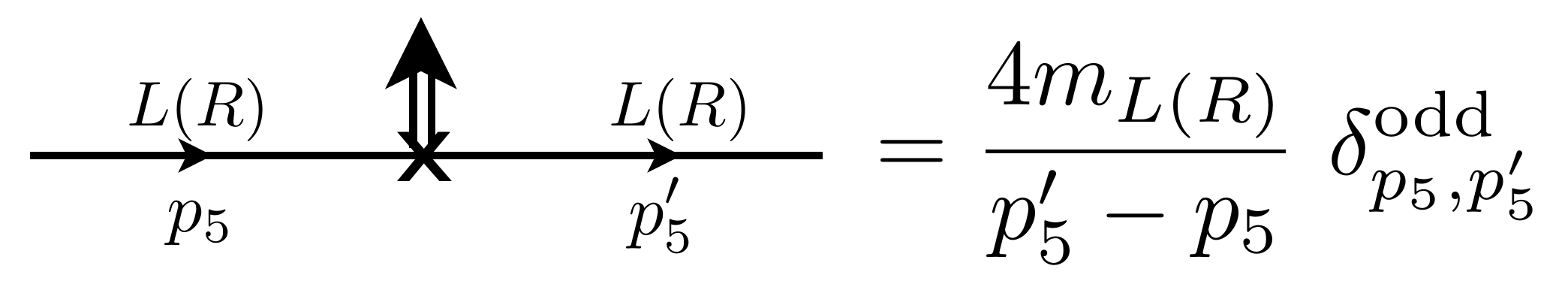},
\end{equation}
where 
\begin{equation}
\delta^\text{odd}_{p_5,p'_5} \equiv \left\{ \begin{array}{cl} 1 & \text{if } p_5+p'_5 \text{ is an odd multiple of } \pi/L \\
0 & \text{if } p_5+p'_5 \text{ is an even multiple of } \pi/L. \end{array} \right.
\end{equation}
This is the familiar Fourier transform of the square wave function, with period $2L$.

The corrections to the scalar two-point function arise from two diagrams, one with a mass insertion on the fermion with a LH zero mode, the other with an insertion on the one with a RH zero mode.
\begin{equation}
\includegraphics[width=5in]{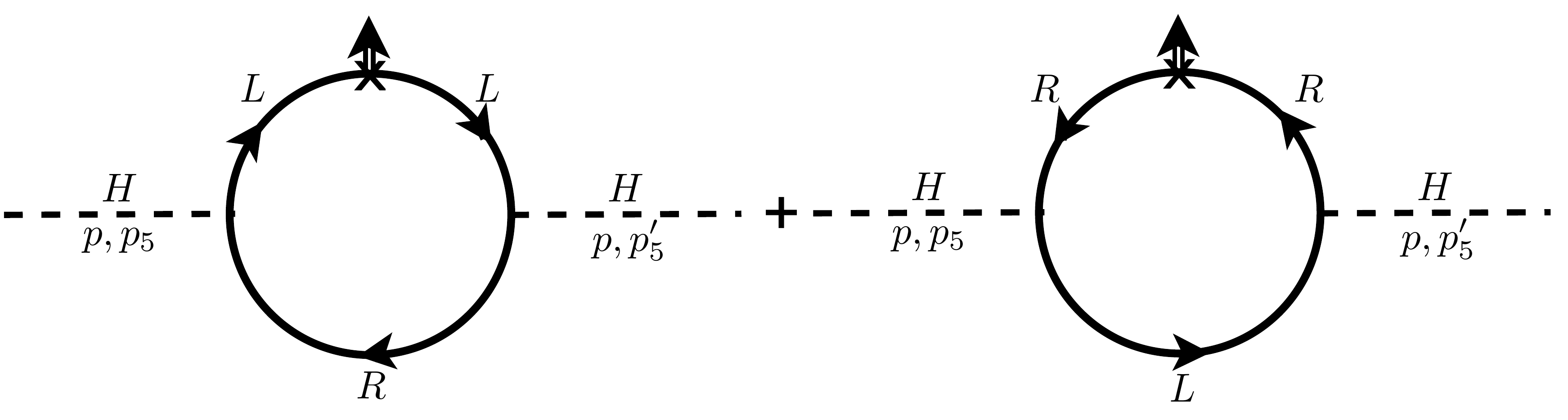}.
\end{equation}
These contributions to the two-point function are linearly divergent:
\begin{equation}
i \Pi_M (0;p_5,p'_5) = i \frac{g^2 \Lambda}{3 \pi^3 \Lambda_0} \left( m_L - m_R  \right) \delta^\text{odd}_{p_5,p'_5} + \text{finite terms}
\end{equation}

Adding a mass insertion diagram to the four-point function only contributes finite terms.

\section{The quantum effective action}

The two- and four-point diagrams we have calculated can now be incorporated into a quantum effective action that is valid at a low scale $\mu$.  We can express this action as follows:
\begin{multline}
S_\text{effective} = \int d^4 x \int^L_{-L} dz \left[   \bar{\Psi}_L( i\displaystyle{\not} \partial-M_L(z) ) \Psi_L  +  \bar{\Psi}_R (i \displaystyle{\not} \partial-M_R(z) ) \Psi_R  + \frac{g}{\sqrt{\Lambda_0}} H \bar{\Psi}_L \Psi_R  + \text{h.c.}    \right.  \\
+ Z_H  \partial_M H  \partial^M H^\dagger   - \left. \left( \Lambda_0^2 + \delta M^2 \right)  |H|^2  -   \frac{\lambda}{4 \Lambda_0}  |H|^4 \right]  \\
 - \int d^4 x~\left[ m^2_{0}~|H(z=0) |^2 +m^2_{L}~|H(z=L) |^2 \right].   \\
 \label{eq:effact}
\end{multline}
To map between our correlation functions and the terms in this effective action, we first note that each amplitude can be written in terms of projection operators $E_{p_5,p'_5} \equiv L \left( \delta_{p_5,p'_5} + \delta_{p_5,-p'_5}  \right)$ acting on ``sub-amplitudes."  The projection operators are the expression for dynamical external scalar legs when the scalar is even under the orbifolding procedure, $H(z) = H(-z)$.  The sub-amplitudes represent Feynman rules arising from bulk and brane localized terms in the effective 5D action.

For the bulk contributions to the two-point function, we have
\begin{align}
i \Pi (p^2;p_5,p'_5) &= E_{p_5,p'_5} i \tilde{\Pi} (P^2) \nonumber \\
&= \frac{1}{2L} \sum_{q_5}  E_{p_5,q_5} E_{q_5,p'_5}   i \tilde{\Pi} (Q^2).
\end{align}
The contribution arising from the bulk mass insertion diagrams is
\begin{equation}
i \Pi_M (0;p_5,p'_5) = i \tilde{\Pi}_M \delta^{\text{odd}}_{p_5,p'_5} =   i \tilde{\Pi}_M \left( \frac{1}{2L}\right)^2 \sum_{q_5,q'_5} E_{p_5,q_5} E_{p'_5,q'_5} \delta^{\text{odd}}_{q_5,q'_5}.
\end{equation}
We can identify $Z_H \equiv \tilde{\Pi}' (Q^2 = 0)$, and $\delta M^2 \equiv -\tilde{\Pi} ( Q^2 = 0)$.  The mass insertion diagrams need to be Fourier transformed back into position space.  We use the identities
\begin{align}
\sum_{p_5~\text{odd}} e^{i p_5 z} &= L \sum_N (-1)^N \delta( z- N L) \nonumber \\
\sum_{p_5~\text{even}} e^{i p_5 z} &= L \sum_N \delta( z- N L)
\end{align}
where the sum over $N$ spans all integers.  The Fourier transform thus corresponds to opposite sign $\delta$-functions on the two branes, $\delta_{q_5,q'_5}^\text{odd} \rightarrow \frac{1}{2} \left[ \delta(z) - \delta(z-L)\right]$.  The brane localized mass terms are then $m_0^2 = -m_L^2 = - \tilde{\Pi}_M/2$.  Finally, the four-point function can be expressed as
\begin{align}
i V_4 (0;p_5,p'_5,p''_5,p'''_5) &= i \frac{\tilde{V}_4}{8} \sum_{\pm} \delta_{0,p_5\pm p'_5 \pm p''_5 \pm p'''_5} \nonumber \\
&= i \left( \frac{1}{2L} \right)^4 \sum_{q_5,q'_5,q''_5,q'''_5} E_{p_5,q_5} E_{p'_5,q'_5} E_{p''_5,q''_5} E_{p'''_5,q'''_5}  \tilde{V}_4~\delta_{0,q_5+ q'_5 + q''_5 + q'''_5}.
\end{align}
and we make the identification $\tilde{V}_4 = \frac{\lambda}{\Lambda_0}$.

In summary, the effective action can be expressed as a function of the UV parameters as in Eq.~(\ref{eq:effact}) with coefficients given by
\begin{align}
Z_H    &= \frac{N_c g^2}{10 \pi^3}\frac{\Lambda}{\Lambda_0}   \nonumber \\
\delta M^2                      &= - \frac{N_c g^2}{18 \pi^3} \frac{\Lambda^3}{\Lambda_0}  \nonumber  \\
\lambda                &=  \frac{N_c g^4}{3 \pi^3} \frac{\Lambda}{\Lambda_0} \nonumber \\
m^2_0  =- m^2_L  & = \frac{N_c g^2}{6 \pi^3} \frac{\Lambda}{\Lambda_0} (m_R - m_L).
\end{align}
We now associate the regulator cutoff $\Lambda$ with the physical scale $\Lambda_0$.  By defining the coupling constants such that they are dimensionless, with the physical scale explicitly appearing in the interaction terms, the quantum corrections (with the exception of the bulk mass term) are all seen to be independent of the scale $\Lambda_0$.

It is interesting that the scalar mass$^2$ receives brane localized contributions of opposite sign on either brane.  This is a severe violation of KK parity.  If this parity were preserved, the two brane localized terms are expected to be identical.  However, the fermion mass terms explicitly violate KK parity.  Quantum effects transmit this breaking of KK parity to the scalar sector in the form of these linear divergences.  

These opposite sign, one loop, brane localized terms vanish, however, when the fermion masses are taken to be identical.  In this scenario, for positive bulk masses, the LH zero mode is localized on the $z=0$ brane, whereas the RH zero mode is localized on the $z=L$ brane.  If the masses are equal, then the profiles are mirror images of each other, and an ``accidental" approximate KK parity is introduced.   

We now choose a convenient normalization for the 5D fields.  We choose a canonical 5D scalar kinetic term, obtained by redefining $H \rightarrow H/\sqrt{Z_H}$,
\begin{align}
S = \int d^4 x &\int^L_{-L} dz \left[ \bar{\Psi}_L \left( i\displaystyle{\not} \partial -M_L (z) \right) \Psi_L   +  \bar{\Psi}_R \left( i \displaystyle{\not} \partial -M_R (z)  \right) t_R   + \frac{\tilde{g}}{\sqrt{\Lambda_0}}     H \bar{\Psi}_L \Psi_R  + \text{h.c.}     \right.  \nonumber \\
  &\left. +\partial_M H \partial^M H^\dagger    -   \tilde{m}^2 |H|^2  -   \frac{\tilde{\lambda}}{4 \Lambda_0}  |H|^4 \right]  - \int d^4 x  \left[  \tilde{m}^2_{0}  \left. |H|^2 \right|_{z=0} +  \tilde{m}^2_{L}  \left. |H|^2 \right|_{z=L} \right].   
\end{align}
The terms in this 5D effective theory are
\begin{align}
\tilde{g}^2                      &= \frac{ 10 \pi^3}{N_c} \nonumber \\
\tilde{m}^2                      &=   \left( \frac{10 \pi^3 }{ N_c g^2}- \frac{5}{9} \right) \Lambda_0^2  \nonumber   \\
\tilde{\lambda}                &= \frac{100 \pi^3}{3 N_c}  \nonumber   \\
\tilde{m}^2_0  =-\tilde{m}^2_L      & =    \frac{5}{3} (m_R - m_L).   
\label{eq:finalresult}
\end{align}
Above, we have assumed $\Lambda \gg \mu$, where $\Lambda$ is the scale that our original Lagrangian with the four-fermion operator was defined, and $\mu$ is the low scale at which we evaluate our 5D effective action.  

There are also finite non-local contributions that arise from quantum corrections.  We have neglected these, as they are typically sub-dominant, and do not have an interpretation as terms which are local in the extra dimensional coordinate.

We note that there are no brane localized quadratic divergences at one loop.  Such terms might  have been expected from considerations of the field content.  In the fermion bubble approximation, brane localized terms arise only from diagrams with insertions of the 5D fermion mass, whose profile explicitly violates translation invariance.

In the presence of fermion bulk masses, the conditions under which the chiral symmetry of the low-energy theory is broken are modified.  In the absence of the boundary terms, the scalar bound states condense for $g^2 > 18 \pi^3/N_c$.  However, the brane localized mass terms can drive condensation as well.  In the next section we explore the conditions for generation of a chiral symmetry breaking condensate, and the resulting spectrum of the theory.


\section{Vacuum Solution and Mass Spectrum}

We have now shown that the low-energy effective theory is one with an additional 5D composite scalar degree of freedom.  The equations of motion and the boundary conditions for this scalar field can be derived from the effective action that we have calculated.  These determine the spectrum of the theory.  

At the high scale, the 5D scalar Higgs field is equivalent to the fermion bilinear $H(z,x) = \bar{\psi}_L (z,x) \psi_R (z,x)$.  With the fermionic orbifold assignments we have made, the orbifold parity transformation of the composite field is
\beq
H (-z) =  \bar{\psi}_L (-z)  \psi_R (-z) = (- \bar{\psi}_L (-z) \gamma^5)(  - \gamma^5\psi_R (-z) ) =  \bar{\psi}_L (z)  \psi_R (z) = H(z).
\eeq
The scalar field is thus orbifold even, which means that when deriving the equation of motion for $H$, we cannot require that the variation itself vanish on the branes.  Rather, the Higgs field is sensitive to the brane localized mass terms.

In this model, chiral symmetry breaking can occur in one of two ways.  First, the coupling constant associated with the four-fermion operator may be sufficiently large that the bulk mass term is driven negative, destabilizing the origin as a vacuum solution.  The bulk quartic coupling then sets the value for the scalar vacuum expectation value.

The other possibility is that the scalar bulk mass$^2$ remains positive, but a negative brane localized mass term pushes the field value away from the origin.  In this case, it is still the bulk quartic coupling that stabilizes the vacuum field solution away from the origin, since we have shown that no brane localized quartic coupling is induced.

The second solution is more interesting, as it distinguishes the behavior of the compact 5D model from the non-compact one.  Unlike the scalar bulk mass, the brane localized terms are sensitive to the values of the fermion bulk mass terms (and thus the relative localization of the fermion zero modes).  Whether chiral symmetry breaking occurs in the extra dimensional model is thus a function of the free parameters of the model.

We now consider solutions to the composite scalar equations of motion.  In the bulk, the vacuum equation for $\langle H(z,x) \rangle \equiv v(z)/(2 \sqrt{ L})$ is given by:
\begin{equation}
v''(z) = \tilde{m}^2 v(z) + \frac{\tilde{\lambda}}{8 \Lambda_0 L} v^3(z).
\end{equation}
This differential equation can be solved in terms of a Jacobi elliptic function, $\text{sc} (x | m)$.  The expression for the vacuum expecation value (vev) is
\begin{equation}
v(z) = \sqrt{\frac{8  \Lambda_0 L \kappa_-}{\tilde{\lambda}}} \text{sc} \left( \left. | z - z_0 | \sqrt{\frac{\kappa_+}{2}} \right| 1- \frac{\kappa_-}{\kappa_+} \right),
\end{equation}
where we have introduced the dimensionless quantities $\kappa_\pm = \tilde{m} ^2 \pm \sqrt{ \tilde{m}^4- \frac{\tilde{\lambda} \tilde{m}^2 v_0^2}{4 \Lambda_0 L}}$.  The quantities $z_0$ and $v_0$ are determined by imposing the boundary conditions.  In order for the low-energy chiral symmetry to be broken, the vacuum energy for the scalar field must be minimized at a non-trivial value for $v_0$.

The only brane localized terms which survive in the large cutoff limit are scalar mass terms proportional to the difference in bulk fermion masses.   These are shown in Eq.~(\ref{eq:finalresult}).  These mass terms, $\tilde{m}_0^2$ and $\tilde{m}_L^2$, set the boundary conditions for the scalar vev equation:
\begin{equation}
\left. \frac{v'(z)}{v(z)} \right|_{z=0} = \frac{1}{2} \tilde{m}_0^2 ~~~~~~~~~\left. \frac{v'(z)}{v(z)} \right|_{z=L}= -\frac{1}{2} \tilde{m}_L^2.
\end{equation}

We can analytically determine the phase boundary by expanding the solution about small $v_0$.  The result is $v(z) \approx v_0 \sinh( | z - z_0 | \tilde{m} )$, and the boundary conditions are then:
\begin{align}
&\left.\frac{v'(z)}{v(z)} \right|_{z=0} = \tilde{m} \coth ( |z_0| m ) = \frac{5}{6} (m_R - m_L) \nonumber \\
&\left.\frac{v'(z)}{v(z)} \right|_{z=L}  = \tilde{m} \coth ( |L-z_0| m ) = \frac{5}{6} (m_R - m_L).
\end{align}
These are satisfied for $z_0 \rightarrow -\infty$, and for $\tilde{m} = \frac{5}{6} (m_R-m_L)$.  We can express this phase boundary in terms of the original four-Fermi coupling $g$, which determines $\tilde{m}$ in the low-energy theory.  The critical coupling is found to be:
\begin{equation}
g_\text{crit}^2 = \frac{18 \pi^3 }{N_c} \left[ 1 + \frac{5}{4} \frac{ (m_R-m_L)^2 }{\Lambda_0^2} \right]^{-1}.
\end{equation}

We now scan the parameter space of the model.  For these purposes, we presume that the fermions are the 5D analogs of the LH third generation doublet and the RH top quark.  In this case, the scalar field then carries the $SU(2)_L \times U(1)_Y$ quantum numbers of a SM Higgs, and when $H$ obtains a vev, the $W$ and $Z$ bosons become massive.  We identify the region of parameter space in which we obtain the correct $W$-boson and top quark masses.

The $W$ mass is well approximated by assuming a flat profile for the lightest $W$-boson mode, and convoluting the flat profile with the vev$^2$:
\begin{equation}
m_W^2 = \frac{g_2^2}{4} \left[ \left(\frac{1}{2L}\right) \int_{-L}^L dz~v(z)^2 \right],
\end{equation}
where $g_2$ is the $SU(2)_L$ gauge coupling of the SM.  The top quark mass is approximated from the Yukawa interaction:
\begin{equation}
m_\text{top} = \frac{\tilde{g}}{\sqrt{\Lambda_0 L}} (N_R N_L L ) \left[ \frac{1}{2L} \int_{-L}^L dz v(z) e^{(m_R - m_L) |z|}\right]
\end{equation}
where $N_{R(L)}$ are the normalization factors for the fermion zero mode profiles, $\Psi_L (z) = N_L e^{-m_L |z|}$, and $\Psi_R (z) = N_R e^{m_R |z| }$.  Note that the $W$ mass depends only the difference between the fermion bulk mass terms (through the effective Higgs potential), while the top quark mass has a quite different dependence arising from the fermion normalization parameters.  The $W$ and top quark masses are thus independently adjustable.

\begin{figure}[t]
\center{\includegraphics[width=4in]{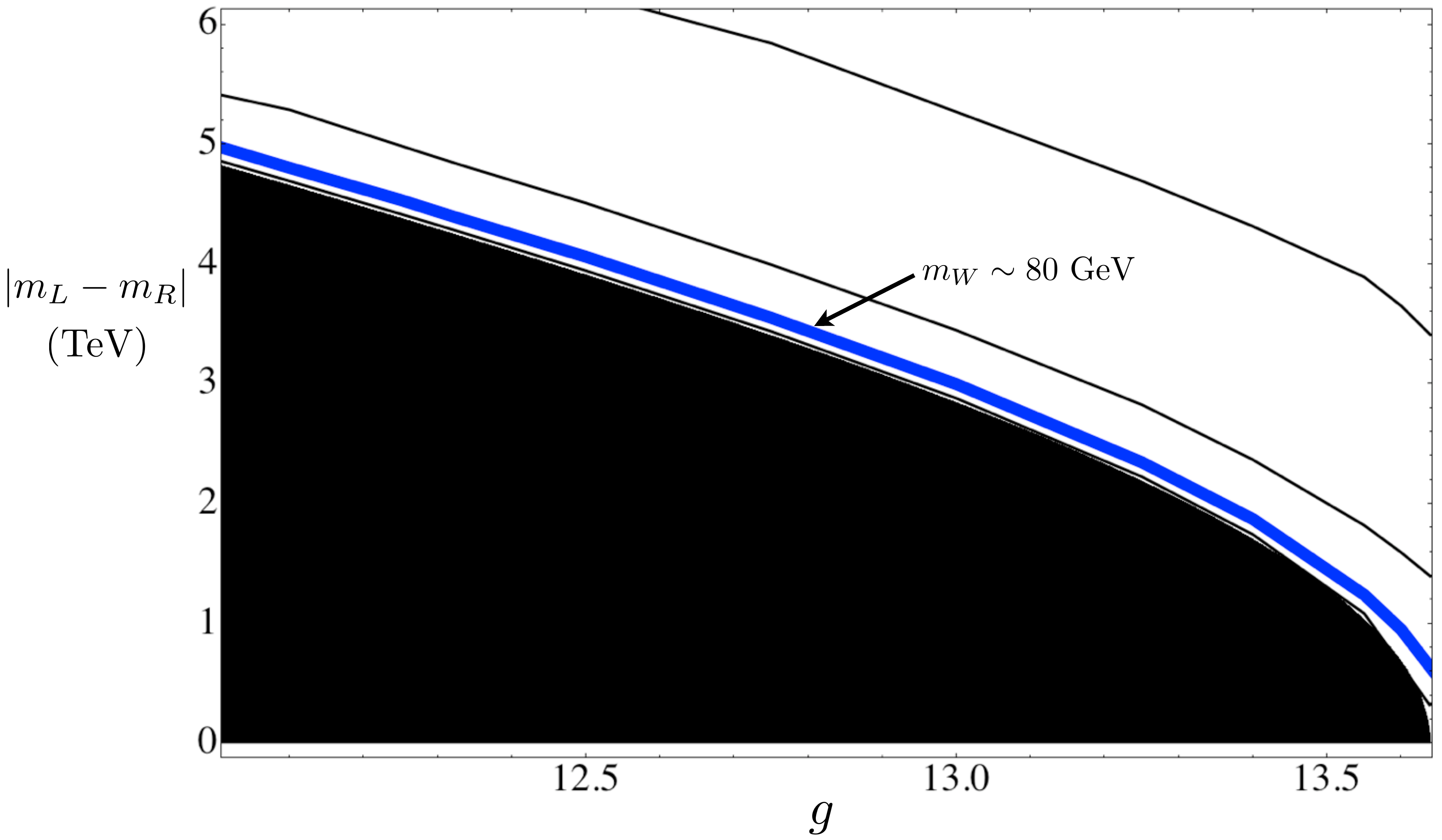}}
\caption{The phase boundary of the model is shown, as a function of $|m_L - m_R|$ and the four-fermion coupling $g$.  The size of the extra dimension is $L = 1$ TeV$^{-1}$, and $N_c = 3$.  Thin solid black lines indicate contours (moving outwards) of $m_W \sim 40,~160,~320$ GeV, while the thick blue line corresponds to $m_W \sim 80$ GeV.}
\label{fig:phaseplot}
\end{figure}

The phase boundary is shown in Figure~\ref{fig:phaseplot} along with contours of $m_W$ as a function of the original four-fermion coupling $g$ and the difference between the fermion bulk mass parameters $|m_R-m_L|$.  We have set the other free parameters to $N_c=3,\text{ and }\Lambda_0 L = 10$.  In Table~\ref{tab:spectrum}, values of $m_L$, $m_R$ and $L$ which give the correct top and $W$ mass are shown, along with the associated value for the Higgs mass.  Additionally, we quote the value of $g^2/g_\text{crit}^2-1$, a rough measure of the fine tuning necessary in the four-fermion coupling to achieve the correct $W$-mass.

We see that for the choice $\Lambda_0 L = 10$, $N_c = 3$, the Higgs is very massive.  In fact, it is above the perturbative unitarity bound.  This can be alleviated by increasing $\Lambda_0 L$, although the fermion ladder approximation begins to break down as $\Lambda_0$ approaches the scale at which the 5D gauge interactions become strong (about $\Lambda L \sim 30$).

\begin{table}[h]
\caption{Choices of the fermion bulk mass parameters that reproduce the SM values for $m_W$ and $m_\text{top}$, and their associated predictions for the Higgs mass.  In the third column, we give a rough measure of the fine tuning necessary to achieve the weak scale from the 5D four-fermion interaction.   All dimensionful parameters are given in units of TeV.  We have set the other free parameters to $N_c=3,~\Lambda_0 L = 10$.}
\begin{center}
\begin{tabular}{|c|c|c|c||c|}
\hline
	$m_L$ &   $m_R$  &  $L^{-1}$      &   $g^2/g_\text{crit}^2-1$ & $m_\text{Higgs}$          \\
	\hline
          9.1            &         18.8                    &        2    &      0.0035 &    1.4                         \\
          \hline	
          9.2          &           17.1                          &        2     &    0.0031    &       1.25                 \\
          \hline
          9.45             &         15.2                 &        2        &    0.0025 &   1.1           \\
          \hline
          10.            &        12.7                     &     2             &    0.0016 &     0.85     \\
          \hline
	\hline
           4.5            &         9.5                      &        1   &       0.014  &     1.4          \\
	\hline
          4.6              &             8.7                  &         1   &      0.012 &         1.3      \\
	\hline
           4.7             &                 7.7            &          1   &        0.010 &        1.1            \\
	\hline
         5.0               &           6.5                    &           1  &     0.006  &      0.85           \\
         \hline
	\hline
           2.25             &                 5.0              &        0.5   &        0.06 &           1.4      \\
	\hline
           2.3             &                4.5               &        0.5   &         0.053 &          1.3       \\
	\hline
           2.25           &             5.4                 &     0.5         &       0.045 &       1.1      \\
	\hline
           2.4           &              3.4                &        0.5       &      0.030 &        0.9    \\
           \hline
\end{tabular}
\end{center}
\label{tab:spectrum}
\end{table}

\section{Conclusions}
\label{sec:conclusions}

We have considered a compactified 5D version of a Nambu--Jona-Lasinio model.  The model is studied by computing quantum corrections to a 5D Yukawa theory in which there are two species of fermions, each with a fermionic zero mode in the spectrum with opposite chiralities.  The scalar field is interpreted as a bound state of the two fermion species.  The classical 4D effective theory at low energies exhibits a chiral symmetry.  Supplementation of the model by a 5D UV composite boundary condition renders the model equivalent at the high scale to one with a 5D bulk four-fermion operator.  The quantum corrections to the low-energy Yukawa model are equivalent to a re-summation of fermion bubble diagrams in the fermion four-point function arising from the four-fermion interaction.

Both bulk and brane localized divergences are generated, although the brane localized divergences are softer than might have been expected.  An accidental remnant of 5D translation invariance on the parent $S_1$ space survives, and protects against one-loop quadratically divergent contributions to the scalar mass$^2$ terms on the branes.  In the presence of fermion bulk mass terms which explicitly violate translation invariance, linear divergences are generated.  Under certain conditions, when the four-fermion coupling exceeds a critical value, these brane localized terms destabilize the scalar vacuum, and drive spontaneous chiral symmetry breaking.

If a portion of the chiral symmetry is weakly gauged, it is expected that this symmetry will be spontaneously broken, as in top condensation models.  We numerically studied such a model, showing that it is possible to realize simultaneously the correct top quark and $W$-boson masses.  This can be seen as an explicit 5D realization of top seesaw models, a deconstructed version of which was studied in~\cite{Cheng:2001nh,He:2001fz}.  The Higgs mass is generically quite large in these models due to the large quartic coupling, likely in conflict with perturbative unitarity and/or electroweak precision constraints.  A more realistic model implemented in warped space may alleviate both of these tensions.
\newpage

\appendix

\noindent {\bf \huge Appendices}

\section{5D Hard Cutoff}
\label{app:euler-maclaurin}
There are many ways in which to implement a hard cutoff in 5D theories, although most do not preserve 5D Lorentz invariance.   For example, a common procedure is to write 5D propagators in mixed position/momentum space, where the propagators are functions of 4D momenta, and of the extra dimensional coordinate, $z$.  It is not practical however, to implement a short distance cutoff in a manner which respects local 5D Lorentz invariance since the extra dimension has been singled out.  Another common approach is to work in a KK-basis, and for each KK mode to integrate over a four-sphere in the 4D momenta.  However, the region in full 5D momentum space that is integrated/summed over is not invariant under the 5D Lorentz group.

An ideal regularization procedure respects 5D Lorentz invariance in the UV, with sub-leading terms generated as finite consequences of non-local finite-volume effects.  To obtain such a regulator, we recall that the Euler-Maclaurin formula allows one to express a sum over integers in terms of an integral and additional corrections:
\begin{equation}
\sum_{n=-\infty}^{\infty} f(n) = \int_{-\infty}^\infty dn f(n) + \lim_{a\rightarrow \infty} \left[ \frac{ f(a) + f(-a) }{2} + \sum_j \frac{B_{2j}}{(2j)!}  \left( f^{(2j-1)} (a) - f^{(2j-1)} (-a) \right) \right]
\end{equation}
Where the $B$ coefficients are Bernoulli numbers.
5D loop integrals thus take the form
\begin{align}
\label{eq:loopint}
\frac{1}{2L} \sum_{k_5} \int \frac{d^4k}{(2 \pi)^4} I (k,k_5)  = \int& \frac{d^5 k}{( 2\pi)^5}  I (k,k_5) +  \frac{1}{2L}\lim_{k_5 \rightarrow \infty} \left[ \frac{1}{2} \int \frac{d^4k}{(2 \pi)^4} \left( I(k,k_5) + I (k,-k_5) \right)+ \right. \nonumber \\ 
&\left. \sum_j \frac{B_{2j}}{(2j)!} \int \frac{d^4k}{(2 \pi)^4} \left( I^{(2j-1)} (k,k_5) - I^{(2j-1)} (k,-k_5) \right) \right],
\end{align}
where the derivatives are with respect to the second argument of the integrand.
On the right hand side, to implement a hard cutoff, we wick rotate and then restrict the momentum integration/summation to the interior of a euclidean four-sphere:  $K^2 \equiv k_0^2+k_1^2+k_2^2 +k_3^2+k_5^2 \le \Lambda^2$.  The final expression for any regulated 5D loop is
\begin{align}
\label{eq:loopintreg}
\frac{1}{2L} \sum_{k_5=-\Lambda}^\Lambda &\int_{K^2 \le \Lambda^2} \frac{d^4k_E}{(2 \pi)^4} I (k_E,k_5) = \int_{K^2 \le \Lambda^2} \frac{d^5k_E}{(2 \pi)^5} I (k_E,k_5) +
\nonumber \\ 
& \frac{1}{2L} \lim_{k_5 \rightarrow \Lambda}  \sum_j \frac{B_{2j}}{(2j)!}   \frac{\partial^{(2j-1)}}{\partial k_5^{(2j-1)}} \left( \int_{k_E^2 \le \Lambda^2-k_5^2} \frac{d^4 k_E}{(2\pi)^4} \left( I(k_E,k_5) + I (k_E,-k_5) \right) \right).
\end{align}
The contribution from the second term in Eq.~(\ref{eq:loopint}) vanishes, since the region of integration in 4-momentum vanishes as $k_5 \rightarrow \Lambda$.

\section{Brane Localized Terms}
\label{app:5Dtranslation}

In~\cite{Carena:2002me}, it was stated that brane localized terms are automatically generated in theories with compact extra dimensions.  There are, however, many cases in which such terms are not generated at the one-loop level.  In this appendix, we discuss these, and provide a symmetry argument for why such terms are protected.  For the purposes of this discussion we use the orbifold language, in which the symmetry principle is most clear.  The extra dimensional space thus begins as a circle, parametrized by angle $\theta$, and is reduced to an interval by identifying points $\theta \leftrightarrow -\theta$.

The reason why most theories generate brane localized kinetic terms is that the orbifolding procedure explicitly violates 5D translation invariance.  In the simplest case, fields can be assigned either even or odd parity under the orbifold identification, a manifestation of this breaking.  Quantum effects will then transmit this breaking to other parts of the theory, creating brane localized kinetic terms, mass terms, and interactions.

\begin{figure}[h]
\center{\includegraphics[width=4in]{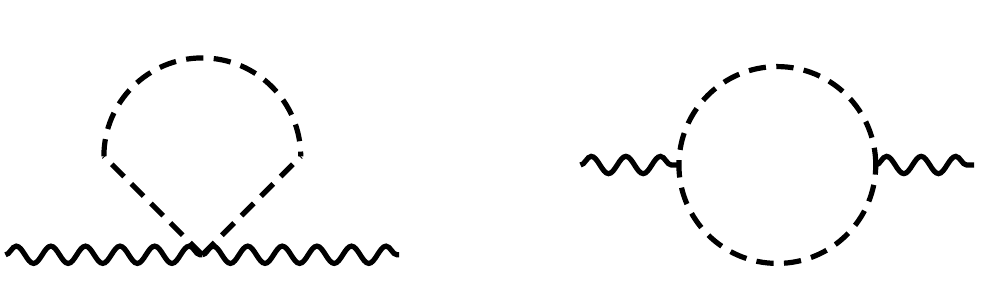}}
\caption{Gauge boson two-point diagrams.}
\label{gaugeloop}
\end{figure}

To see this in action, consider a 5D scalar field with no 5D mass term.   The propagator for a scalar which is even or odd under the orbifold assignment is given by~\cite{Georgi:2000ks}:
\begin{equation}
\Delta(p; p_5,p'_5) = \frac{i}{2} \frac{1}{p^2 - p_5^2} \left\{ \delta_{p_5,p'_5} \pm \delta_{p_5,-p'_5} \right\}
\end{equation}
Now let us add gauge interactions and consider the gauge boson two-point function.  There are two diagrams shown in Figure~\ref{gaugeloop}, although one creates a non-transverse structure which is completely canceled by a portion of the second.  This is a consequence of gauge invariance.

The diagram contains the following numerator structure which arises from the two scalar propagators in the loop:
\begin{align}
\sum_{k_5,k'_5} \frac{1}{D(k_5,k'_5)}& \left( \delta_{k_5,k'_5} \pm \delta_{k_5,-k'_5} \right) \left( \delta_{p_5+k_5,p'_5+k'_5} \pm \delta_{p_5+k_5,-p'_5-k'_5} \right) \rightarrow \nonumber \\
& \sum_{k_5} 
 \frac{1}{D(k_5,k'_5)} \left\{ \delta_{p_5,p'_5} + \delta_{p_5,-p'_5} \pm \delta_{2 k_5,p_5-p'_5} \pm \delta_{2 k_5,p_5+p'_5} \right\}
\end{align}
The last two terms which do not conserve 5D momentum correspond to brane localized terms, and are divergent when the full expression is evaluated.  However, note that they come with either positive or negative coefficient depending on whether the scalar has positive or negative orbifold parity.  This shows that if a theory is constructed which has two such scalars, with opposite orbifold parity and equal gauge coupling, that the brane localized divergences will cancel.  The reason for this is that the enhanced spectrum is identical to that of the theory before the orbifolding has taken place, and therefore has all the field content of the complete circle before orbifolding.  5D translation invariance on the full un-orbifolded circle protects against the generation of brane localized terms.

Now consider a 5D fermion on the same spacetime.  If the bulk mass of the fermion vanishes, the fermion propagator (with the Dirac structure made explicit) is given by
\begin{equation}
\Delta(p; p_5,p'_5)  =  \frac{i}{2} \frac{1}{\dslash{p} + i \gamma^5 p_5}
\left( \begin{array}{cc} 
\bold{1}_{2\times2} \cdot \left(\delta_{p_5,p'_5} \pm \delta_{p_5,-p'_5} \right) & \bold{0}_{2 \times 2} \\
\bold{0}_{2 \times 2}  & \bold{1}_{2\times2} \cdot \left( \delta_{p_5,p'_5} \mp \delta_{p_5,-p'_5} \right) \end{array} \right).
\end{equation}
The fermion propagator contains two parts, one of which is orbifold even, and the other odd.  These correspond to the right- and left-handed components of the 5D Dirac fermion.  As with the case of two scalar fields, these degrees of freedom act together in diagrams, and can potentially conspire to make brane localized terms vanish.  The question of whether or not brane localized terms are generated thus comes down to the interplay of these two parts of the fermion propagator in particular processes.  In the two-point function we calculate for the Yukawa theory, only bulk renormalization takes place, and no brane terms are generated.   In contrast, for the case of anomalies, the components of the propagator work together such that only brane localized divergences are generated~\cite{Georgi:2000ks}.


\section*{Acknowledgements}

We thank Csaba Cs\'aki for useful conversations during the course of this work.  
D.B., J.H., and J.S. thank Cornell University for hospitality during the course of this work.
D.B, J.H., and J.S.\ are supported in part by the DOE under grant number DE-FG02-85ER40237, and in part by the Syracuse University College of Arts and Sciences. P.T.\ is supported in part by the NSF under grant number PHY-0355005, an NSF graduate research fellowship, and a Paul \& Daisy Soros Fellowship for New Americans.


\bibliographystyle{utphys.bst}
\bibliography{doubleCHiggs}

\end{document}